# Strategic AI Governance: Insights from Leading Nations

Dian W. Tjondronegoro

Business Strategy and Innovation, Griffith University, d.tjondronegoro@griffith.edu.au

*Abstract*: Artificial Intelligence (AI) has the potential to revolutionize various sectors, yet its adoption is often hindered by concerns about data privacy, security, and the understanding of AI capabilities. This paper synthesizes AI governance approaches, strategic themes, and enablers and challenges for AI adoption by reviewing national AI strategies from leading nations. The key contribution is the development of an EPIC (Education, Partnership, Infrastructure, Community) framework, which maps AI implementation requirements to fully realize social impacts and public good from successful and sustained AI deployment. Through a multi-perspective content analysis of the latest AI strategy documents, this paper provides a structured comparison of AI governance strategies across nations. The findings offer valuable insights for governments, academics, industries, and communities to enable responsible and trustworthy AI deployments. Future work should focus on incorporating specific requirements for developing countries and applying the strategies to specific AI applications, industries, and the public sector.

## 1 INTRODUCTION

Artificial Intelligence (AI) holds immense potential to transform various sectors, yet its adoption is often slower than technological advancements due to concerns about data privacy, security, and the understanding of AI capabilities. AI's reliance on data for machine learning and smart applications raises significant issues around data privacy and security. Governments worldwide can learn from the rapid AI-enabled business innovations in leading nations and play a pivotal role in developing contextually aware business strategies and governance in the AI era. For instance, AI applications in agribusiness can promote sustainable production through real-time data monitoring, while in healthcare, AI can assist in complex decision-making and enhance the quality of care.

This paper aims to synthesize AI governance approaches, strategic themes, and enablers and challenges for AI adoption by reviewing existing surveys and literature from academic, government, and industry sources. The key contribution is the development of an EPIC (Education, Partnership, Infrastructure, Community) framework, which maps AI implementation requirements to fully realize social impacts and public good from successful and sustained AI deployment. Through a multi-perspective content analysis of the leading nations' latest AI strategy documents, this paper provides a structured comparison of AI governance strategies across nations. The findings offer valuable insights for governments, academics, industries, and communities to enable responsible and trustworthy AI deployments.

The EPIC framework emphasizes the importance of education and training to build AI literacy and skills, fostering partnerships for research and development, establishing robust infrastructure for data and AI, and ensuring community impact through responsible AI innovations. By addressing these areas, the framework aims to guide the development of AI strategies that are ethical, transparent, and beneficial to society.

## 2 LITERATURE REVIEW

Existing reviews on national AI strategies have identified current and emerging themes and similarities across different countries. For example, Bareis [2] analyzed 14 AI policy documents from China, the USA, France, and Germany published between Dec-2017 and Jun-2021 to investigate how to integrate AI tech into the functioning and structures of society. All these nations have established AI as inevitable and massively disrupting technological development, building on the grand legacy and international competition as rhetorical devices. The composition for AI policy narratives is similar, emphasizing leadership intervention and articulation of opportunities and national pathways. However, their objectives and interpretations of AI are different due to the countries' vast cultural, political, and economic differences.

Fatima et al. [4] analyzed 34 national strategic plans published between Oct-2016 and Jan-2020 to understand the prominence of identified key themes by quantifying the number of national strategies covering these themes. The first two key themes are the top priority of public sector functions and industries for AI. The next key theme is data as a critical ingredient for AI systems, enabling the development of algorithms and systems' outputs. AI's performance is largely affected by the data used to train algorithms. AI thrives on large datasets in various forms and from multiple sources. Therefore, the key concerns are data exchange, regulations, privacy, and security. Similarly, Radu [14] analyzed 12 national strategies of South Korea, Canada, Japan, Singapore, Finland, China, UAE, France, the UK, Sweden, Germany, and the US published between April 2016 and Febr-2019. The analysis underlined that the preferred AI policy development recipe is uniting the political will and public resources with the industry interests. Therefore, AI industry growth is desired, enabled, and facilitated by the government but may result in imposed limitations on AI development and implementation, such as ethical issues, making it hard to disentangle public interest policies from market dominance interests.

Schiff [15] analyzed 112 AI ethics documents from public, private, and NGO organizations worldwide published between January 2016 to July 2019 and social media and news collected from the fall of 2018 to early 2020. The analysis showed that all organizations have a common interest in ethical topics and reveal organizational differences in the ethical framing of several key issues, including the perceived scope of responsibility, the role of experts and the public, and the tension between economic and other values. The public sector has stronger attention on economic growth and unemployment, NGOs have a broader range of ethics topics covered, and the private sector has higher relative priority on social responsibility and trust. Zuiderwijk [24] analyzed the implications of AI use in public governance and research agenda from 26 pieces of literature from 2018 to 2020. The analysis identified themes from the potential benefits of AI use in government: and challenges of AI use, namely.

Existing reviews on AI adoptions have identified the motivations, enablers, and barriers within different countries and continents' public and private sectors. Stanford's AI Index Report [23] was based on most up-to-date and comprehensive report on the state of AI R&D, technical performance, technical AI ethics, the economy and education, AI policy and governance. The key takeaways are: 1) US and China dominated cross-country collaborations measured by joint AI publications, 2) Industry-affiliated researchers contributed 71% more publications year-over-year at ethics-focused conferences, 3) nine out of ten state-of-the-art AI systems are trained with big data since 2021, 4) AI-related bills increased significantly in 25 countries, particularly, Spain, UK, and US.

IBM's adoption report [6] was based on a poll conducted from 30 March to 12 April 2022 on a representative sample of 7502 business decision makers with a mix of seniority, 500 in each country from the US, China, India, UAE, South Korea, Australia, Singapore, Canada, UK, Italy, Spain, France, and Germany, and 1000 from Latin America including Brazil, Mexico, Columbia, Argentina, Chile, and Peru. The top themes of AI are automation and skills, building trust, and making businesses sustainable. The analysis studied the key themes, drivers, and barriers for AI adoption. Similarly, Mikalef [11] analyzed survey data collected between October 2020 and January 2021 from municipalities in Europe, including Norway (71%), Germany (22%), and Finland 7%), with target respondents chief digital officers (65%) and higher-level tech managers (35%) in municipalities. The analysis highlighted that the critical determinant factors for fostering AI capabilities, including perceived benefits, organizational factors, and perceived pressure and incentives citizen from government. Furthermore, Vial [20] analyzed interview data of key AI leaders from industry within North American countries and AI professionals working in different contexts. The analysis emphasized that AI developments to bring solutions out of the lab will only succeed if organizations manage data access throughout the development and production life cycle.

Finally, OECD's report [13] examined key factors that determine the AI readiness ranking, focusing on four topics: governance, infrastructure and data, skills and education, government, and public services. In addition, they developed five imperatives for harnessing the power of low-cost prediction based on the economics of AI and called for a unified effort to move forward and reduce the knowledge gap between scientific and interdisciplinary efforts.

Based on these existing reviews, Table 1 summarizes the motivations, indicators of capacity, and enablers of AI adoption, while Table 2 summarizes the political, economic, societal, and technological (PESTEL) analysis of AI barriers and challenges. All these considerations inform AI implementation frameworks of the critical considerations along each process and the success factors, which can serve as a checklist and criteria for milestone completions and compliance audits before, during, and after the implementation of AI.

Table 1. Motivations, Capacity, and Enablers of AI (summarized from [24] [6] [23] [11, 13] [20])

| | |
|---|---|
| **Motivations for AI Adoption** | Benefits: efficiency and performance, risk identification and monitoring, economic, data and information processing, service provision, society at large, decision making, engagement and interaction, and sustainability |
| | Adopting themes include automation and skills, building trust, and sustainable business. The benefits are automated IT, business, or network processes, including cost savings and efficiency, improvements in IT, network performance, or customer experiences. |
| **Indicators for AI Capacity** | Capacity: cross-country collaborations, increased AI ethics research, state-of-the-art AI systems and training data, AI bills. |
| | Capability: perceived benefits of AI, organizational factors (perceived financial costs and organizational innovativeness), perceived citizen pressure, perceived government pressure and incentives, and effect of regulatory guidelines. |
| | Readiness: governance, infrastructure and data, skills and education, government, and public services |
| **Drivers/Enablers** | Data accessibility enablers: address it at the onset, prioritize it as a business issue before considering the technological challenges, and ensure data access throughout the development and production life cycle. |
| | Adoption enablers: AI advancements that make it more accessible, reduce costs and automate key processes, Increase of AI embedded into standard and off-the-shelf business applications, competitive pressure, Demand due to covid-19 pandemic, pressure from consumers, Directives from leadership, Company culture, Labor/skills shortages, Environmental pressure |

Table 2. PESTEL analysis on AI adoption barriers and themes (summarized from [6] [20] and [24])

| | |
|---|---|
| **Political:** | Siloed AI initiatives away from operations limit the impact of AI project |
| | Organizational resistance and management's attitude towards risks and traditional bureaucratic government |
| | AI use that undermines fundamental values due to lack of accountability or transparency of process and decision making |
| **Economical** | The price to implement or adopt is too high |
| | Projects are too complex or difficult to integrate and scale |
| | Too much data complexity (the economy of data). Interpretation of data may lead to information overload |
| **Societal** | Limited AI skills, expertise, or knowledge |
| | Most companies have not taken steps towards trustworthy AI (leading to societal distrust or misconceptions about AI) |
| | Ethical and legitimacy concerns related to moral dilemmas and unethical use of data |
| | Impact of AI on the labor market (automation and skills, building trust, and making business sustainable) |
| **Technological** | Data availability, quality, and structure |
| | Lack of tools or platforms to develop models |
| | Lack of all stakeholder's understanding of the full dimensions of data quality |
| **Environmental** | Ecological and environmental aspects and their effects on how industries (corporate social responsibility and sustainability practices) |

## 3 ANALYSIS OF NATIONAL AI STRATEGIES

A key benefit of learning from national strategies is to see through a third-party view the interests and considerations of AI focusing on the public good, as most governments see themselves as the catalyst, enabler, and regulator for AI developments and deployments. This Section discusses the data collection process, data analysis, and the results from analyzing leading nations' AI strategic documents.

### 3.1 Selecting the leading nations

A combined ranking method is applied for selecting the leading nations based on the top-10 ranked countries from existing multi-factors analysis of AI capacity scoring. Combining the multi-ranking system ensures that the leading nations are determined based on key performance metrics in the last five to ten years and combined from reputable sources. A country gets scored based on the ranking from each source, with 10 being the highest score and 0 being the lowest score if not for ranked within the top 10. For example, rank-1 is scored 10, while rank-10 scored 1. As shown in Table 3, the top-ranked countries included in our study have been included in the top-10 from at least two ranking sources, except The Netherlands, due to their high (88.2%) percentage of internationally collaborative articles (international articles % as reported by Nature index) in the field, higher than that of Italy, Spain, South Korea, Singapore, and India. Table 4 provides the details of the national AI strategy documents (data collected) for analysis, sorted from the oldest to the most recently published. The followings summarize the ranking methodologies from each source.

**Oxford's** overall Government Artificial Intelligence Readiness Index 2018/**2019** used eleven input metrics grouped under four high-level clusters: governance, infrastructure, and data; skills and education; and government and public services. The data is based on desk research and databases such as the number of registered AI start-ups and the UN eGovernment development index indices. URL: https://www.oxfordinsights.com/ai-readiness2019.

**Nature's** Index for top-25 countries or territories in artificial intelligence ranked by total article share in the field from 2015 to **2019**, indicating the percentage of internationally collaborative articles in the field. URL: https://www.nature.com/nature-index/supplements/nature-index-2020-ai/tables/countries.

**Stanford's** AI Vibrancy scores are based on normalized scores from a comprehensive desk review of 2017 to **2021** data on the economy and R&D pillars. These include commercialization (private investment, number of newly funded companies), R&D outputs (total number of journal and conference papers, and repository publications, and the total number of citations from each type of publications), R&D impact (total patent filing and granted, annual h-index), and Skills: relative AI skills, AI talent concentration. URL: https://aiindex.stanford.edu/vibrancy.

**Scimago's** country ranking is determined by the author's origin from publications, calculated as the number of documents, citable documents, citations, self-citations, citations per document, and h-index from 1996-**2021**. URL: https://www.scimagojr.com/countryrank.php?category=1702.

**IBM's** AI Adoption index is based on companies' locations that have deployed or explored AI in **2022**. Their report highlighted that 53% of IT professionals said they had accelerated their rollout of AI in the last 24 months; hence we combined the ranking from deployment and exploring. URL: https://www.ibm.com/watson/resources/ai-adoption.

Table 3. Combined ranking of the leading nations to be included in our analysis (sorted by the combined ranking)

| **Country** | **Oxford Score** | **Nature Score** | **Scimago Score** | **IBM rank Score** | **Stanford Score** | **Total Score** | **Ranking** |
|---|---|---|---|---|---|---|---|
| USA | 7 | 10 | 9 | 2 | 10 | 38 | **1** |
| China | 0 | 7 | 10 | 10 | 9 | 36 | **2** |
| UK | 9 | 9 | 6 | 5 | 6 | 35 | **3** |
| Canada | 5 | 5 | 10 | 5 | 7 | 32 | **4** |
| Germany | 8 | 8 | 5 | 6 | 4 | 31 | **5** |
| India | 0 | 0 | 8 | 8 | 8 | 24 | **6** |
| Singapore | 10 | 0 | 0 | 9 | 1 | 20 | **7** |
| France | 3 | 6 | 4 | 4 | 0 | 17 | **8** |
| Japan | 1 | 3 | 7 | 0 | 0 | 11 | **9** |
| Italy | 0 | 0 | 3 | 7 | 0 | 10 | **10** |
| Spain | 0 | 0 | 2 | 5 | 0 | 7 | **11** |
| Australia | 0 | 2 | 0 | 2 | 3 | 7 | **11** |
| South Korea | 0 | 0 | 0 | 2 | 5 | 7 | **11** |

Table 4. Leading Nations' AI Strategy Documents (sorted from the oldest to the most recent of the last update)

| **Country** | **National AI Strategy Document** | **Issuing Agency** | **Last updated** | **Num of pages** |
|---|---|---|---|---|
| China | Next-Generation Artificial Intelligence Development Plan [22] | CCP Central Committee and the State Council, Republic of China | **Jun-17** | 28 |
| France | For a Meaningful Artificial Intelligence: Towards a French and European Strategy [21] | Cedric Villani and Team, the task assigned by French Prime Minister Edouard Philippe | **Mar-18** | 154 |
| USA | The National AI Research and Development Strategic Plan[19] | National Science and Technology Council, USA | **Jun-19** | 50 |
| Japan | AI for Everyone: People, Industries, Regions, and Governments (Tentative Translation) [9] | Integrated Innovation Strategy Promotion Council Decision Japan | **Jun-19** | 74 |
| South Korea | National Strategy for Artificial Intelligence [10] | Ministry of Science and ICTArtificial Intelligence Policy Division Korea | **Oct-19** | 62 |
| The Netherlands | Strategic Action Plan for Artificial Intelligence [12] | Ministry of Economic Affairs and Climate Policy, the Netherlands | **Oct-19** | 64 |
| Singapore | National AI Strategy: advancing our smart nation journey[16] | Smart Nation Singapore | **Nov-19** | 45 |
| Spain | National Strategy for Artificial Intelligence [17] | Ministry of Science, Innovation and Universities, Spain | **Nov-20** | 89 |
| Germany | Artificial Intelligence Strategy of the German Federal Government [5] | Federal Ministry of Education and Research, the Federal Ministry for Economic Affairs and Energy, and the Federal Ministry of Labour and Social Affairs, Germany | **Dec-20** | 31 |
| India | Responsible AI #AI For all [7] | NITI Aayog (Policy Think Tank of) Government of India | **Feb-21** | 64 |
| Australia | Australia's AI Action Plan [1] | Australian Government Department of Industry, Science, Energy, and Resources | **Jun-21** | 30 |
| Canada | Maximizing Strengths and Spearheading Opportunity [3] | Information and Communications Technology Council, Canada | **Sep-21** | 41 |
| UK | National AI Strategy [18] | The Secretary of State for Digital, Culture, Media, and Sport by Command of Her Majesty, UK | **Sep-21** | 35 |
| Italy | Strategic Programme on AI: 2022-2024 [8] | It was jointly developed by the Ministry of Education, University and Research, the Ministry of Economic Development, and the Minister of Technological Innovation and Digital Transition. | **Nov-21** | 40 |

### 3.2 Analysis of High-Level Summary and Overview

The analysis of executive and high-level summaries from each leading nation's AI strategy has identified the shared values, mission, and objectives for AI implementation.

**Shared values and Missions:** all the leading nations emphasized responsible and public good-oriented development and application of AI systems [Germany] to bring positive benefits to society [Canada] with meaningful AI [France], ensuring AI for all/everyone [India, Japan], capitalizing on AI's societal and economic opportunities, safeguarding the public interests of AI for prosperity and well-being [Netherlands]; remarked by the deployment of scalable, impactful AI solutions in key sectors of high-value and relevance to citizens and businesses [Singapore], while adopting trusted, secure, and responsible AI [Australia]. AI ecosystem is the digital and societal infrastructure, including processes, policy, and skills. The environment for AI is collective across all stakeholders with different interests, concerns, and priorities. Therefore, national capacity for enabling the AI ecosystem and increasing the efficacy of AI solutions is essential to realize intelligent systems' socioeconomic and environmental benefits.

The top nations in AI have the imperatives (i.e., government pressure) to maintain their leadership in the field. For example, the president's executive order for a concerted effort to promote and protect AI technology and innovation in the country to maintain American leadership in AI [USA]. The next-generation AI development plan aims to accelerate the construction of an innovative nation and global power in science and technology, seizing the major strategic opportunity of AI developments, which builds on the "first-mover" advantage [China]. The key ambition is to maintain a global AI superpower by leading the world through a research and innovation powerhouse, a hive of global talent, and a progressive regulatory and business environment [UK].

**Objectives**: All the leading nations' AI strategies emphasize the role of government as the catalyst to coordinate the efforts of government, industry, academics, and society in increasing the efficacy of AI solutions to be adopted. The government plays a critical role in improving the national capacity and its international standing in enabling the AI ecosystem, including the digital and societal infrastructure, processes, policy, and skills, which are essential for AI adoption. While each country has specific objectives for enabling the AI ecosystem, all countries placed data as equally important as AI, noting that combining data and AI capabilities will directly contribute to the country's vision and mission.

The followings will discuss each of the country's specific focus to provide a view of the differences and similarities of the leading nation's goals and strategies for AI (listed in alphabetical order).

**Australia**: the overview of the AI Action Plan mentions data four times within the focus area of a global leader in developing and adopting trusted, secure, and responsible AI. The key considerations are: 1) privacy, 2) safe and transparent public sector data, 3) effective, safe, and secure use of data in government and policy setting, and 4) consumer rights.

**Canada**: the foreword from the Executive Director of the Pan-Canadian AI strategy states that AI will continue to play an important role in understanding the world by finding patterns in data realms. During turbulent times due to geopolitical, economic, and climate instability, the shared value is deploying AI to bring positive benefits to society.

**China**: the statement on its strategic situation emphasizes China's unique advantage in AI development, namely, the accumulation of technological capabilities and massive data resources, as well as the organization integration of both the huge demand for applications and the open market environment.

**France**: the executive summary highlights the importance of data within all five key parts of the strategy for meaningful AI, namely: 1) building a data-focused economic policy, 2) promoting agile and enabling research, 3) working for a more ecological economy, 4) ethical considerations of AI, and 5) inclusive and diverse AI.

**Germany**: data is an integral part of AI strategy priorities, particularly in: 1) bolstering AI expertise, 2) strengthening national research structures, 3) AI research for healthcare, environment and climate, aerospace, and mobility applications, 4) Networking and international cooperation 5) norms, standardization, and test spaces for innovations, 6) AI-enabled government services, such as healthcare and long term care, public administration, 7) regulation in work settings and for product safety, 8) society data sharing and skills.

**India**: the "AI for all" mission promotes responsible AI system considerations for data, including 1) safe and reliable deployment, 2) explainability, 3) consistency against biases across stakeholders, 4) inclusion of services, 5) decisions accountability, 6) privacy, and 7) security.

**Italy**: the summary of key policies emphasizes launching Italian AI research data and software platforms to create government applications. Their key objectives include 1) integrated datasets for open data and open AI models, 2) a common Italian language dataset for AI development, 3) datasets for AI-based analytics feedback 4) datasets for service improvement in public applications.

**Japan**: the key goals of "AI for everyone" include: 1) education reform, reconstruction of the research and development system, 2) social implementation, 3) development of data-related infrastructure, 4) digital government, and 5) ethics.

**South Korea**: the presidential committee on the fourth industrial revolution selected data, network, and AI as three new innovative industries, and data is prominent within two key agendas: 1) AI infrastructure enhancement and 2) diffusing AI tech across all industry areas.

**Singapore**: to become a leader in developing and deploying scalable and impactful AI solutions in key sectors of high value and relevance to our citizens and businesses, the two key components are 1) national AI projects, 2) AI ecosystem enablers, including Triple helix partnership, AI talent and education, data architecture, progressive and trusted environment, and international collaboration.

**Spain**: the president's prologue states that the widespread data use and management through algorithms and autonomous systems have many ethical and moral implications. The key measures emphasize: 1) data solutions through fostering scientific research, technological development, and innovation of AI, 2) establishment of a data office and appointment of a chief data officer to establish a data platform and technological infrastructure to support AI.

**The Netherlands**: the executive summary states that the country is well-placed to capitalize on AI's societal and economic opportunities and safeguard the public interests of AI for prosperity and well-being. The key enablers are 1) world-class networks, data centers/hosting providers, 2) a digitally active and skilled population, 3) a research community that does ground-breaking research on applied AI, 4) collaboration in the field of talent and knowledge development, ensuring access to knowledge, data sharing, which contribute to increasing societal acceptance of AI, promotion of AI within new economic activity by playing a vital role towards application areas/sectors.

**UK**: the executive summary states the key drivers for progress, discovery, and strategic advantage in AI are: 1) access to people, data, computing, and finance, 2) planning and investment for the long-term needs of data and AI ecosystem, 3) the appropriate national and international governance of data and AI technologies.

**USA**: The latest national R&D strategic plan establishes a set of objectives for federally funded AI research. Advancing data-focused methodologies for knowledge discovery is a key aspect of its first objective of making long-term investments in AI research. The fifth key objective is to develop shared public datasets and environments for AI training and testing.

### 3.3 Analysis of Key Themes and Concepts Clusters

For content analysis and identification of the key concepts, themes, and passages within the collection of national strategic documents, *Leximancer* was used as a text analytics tool to produce a conceptual map that provides a bird's eye view, representing the main concepts and their relationships. The tool helped identify the **top themes** of the full documents of all the leading nations' AI strategy documents. The concepts were identified automatically with no limit to reveal diverse topics. The compound concepts were set based on manual discoveries after ten initial runs: "AI and algorithms," "data and based," "information or data," "public and policy and innovation," "R&D or (research and development)," "system or systems," and "AI or Artificial Intelligence." By combining the top five concepts for each nation, we identified the top-60 concepts. Then, they were used for the visualized analysis of the top four themes and concepts as depicted in Figure 1. It demonstrates **"data" is a central theme** that overlaps with all other themes. Hence, by analyzing the documents with data as the central theme, we can do a focused analysis that overlaps with all other key topics.

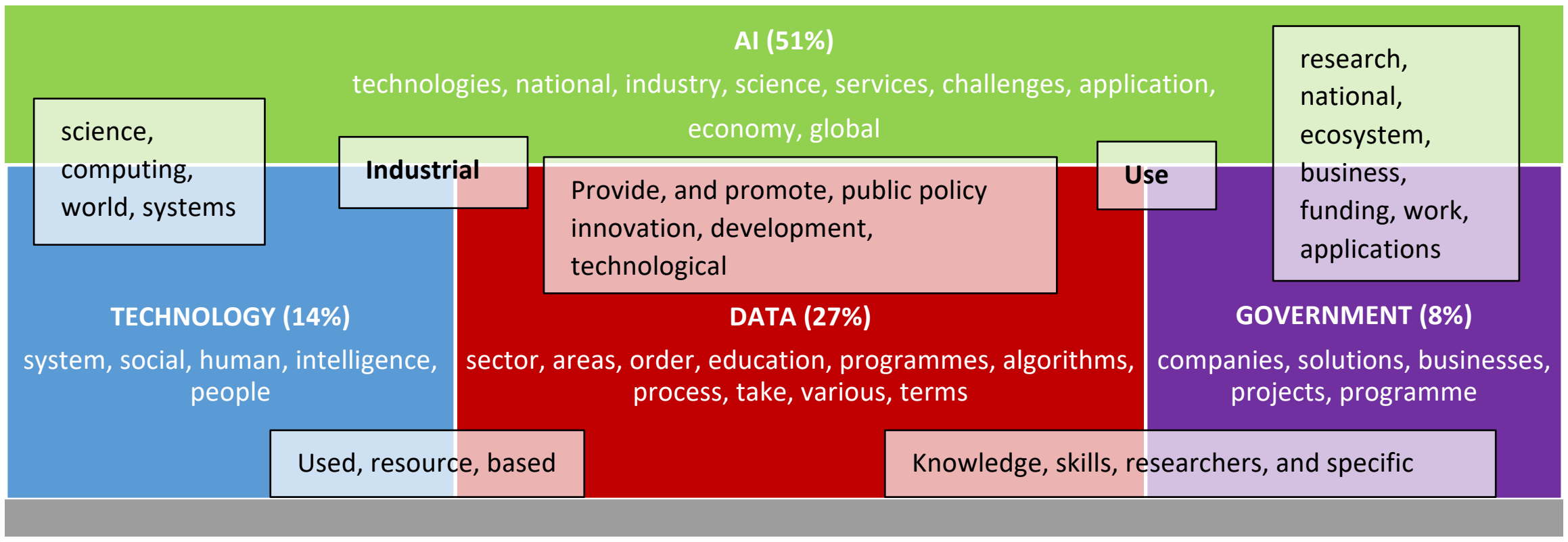

Figure 1. Top four themes based on Leximancer (each theme's ratio in % is based on the total of 11,246 hits/occurrence)

"**Industrial**" is a concept that overlaps between AI, technology, and data, which can be interpreted as industrial applications of AI needs data and technology. Similarly, "**use**" is the concept that overlaps between AI, data, and government, which can be interpreted as the use of data in AI being regulated by the government. Data overlap with technology in terms of being interdependent for "**used, based, and resource**"—for example, data-based technology. Data overlap with the government in terms of "**knowledge, skills, researchers, specific**," as the government has a significant role in supporting training of specific skills and knowledge for data researchers. AI and technology overlap in terms of the "**science, computing, systems, and the world**" as the technological science and systems of AI must be world-class to be competitive and reliable. Finally, AI and government overlap in terms of the "**research, national, ecosystem, business, funding, work, applications**," as the government must support the funding, ecosystem, and work that supports the national research ecosystem and applications.

Leximancer identified the top five hits, whereby data that overlaps with the related concepts: 1) educative policies for gender equality, 2) fostering technology capacities in strategic digital value chains and enhanced cybersecurity, 3) social welfare and sustainability underpin AI deployment, 4) networks and collaborations, 5) AI training for interdisciplinarity.

### 3.4 Data-focused Strategies for AI

We identified the recommendations (best practice, patterns) and things to address (issues and anti-patterns) by analyzing Leximancer's identified favorable and unfavorable sentiment-analyzed passages about data - passages are the set of sentences where data and sentiment-related concepts co-occur. We focused on extracting each passage into a list of actions or strategies and clustering them into thematic paragraphs before grouping them into key headings. Figure 2 depicts the distribution of the data-focused strategies in terms of which countries covered them and how many passages from the national AI documents were analyzed to identify these strategies. This section will list the recommendations while referencing the source documents, noted by the countries instead of the citation.

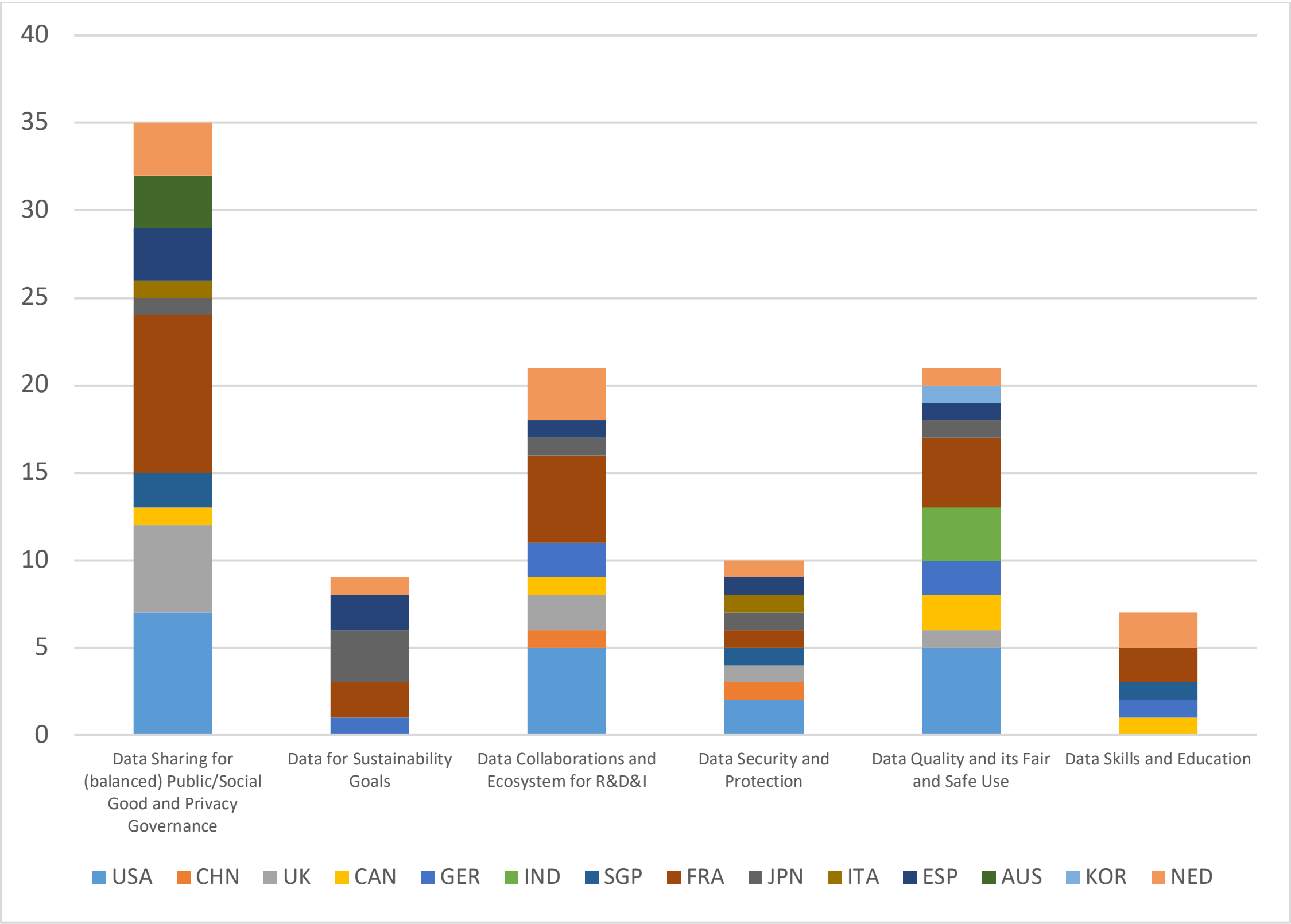

Figure 2. Distribution of data strategies based on themes and countries

*3.4.1. Data Sharing for (balanced) Public/Social Good and Privacy Governance*

**Theme-1: Make data a common good** whereby the government acts as a trusted third party for public authorities to introduce new ways of producing, sharing, and governing data and encouraging economic players to share and pool their data, and supporting public auditing by incentivizing the public to make data available for research [France, USA, Singapore, UK, Australia]. •Review the law to enable "the data of public interest" as a form of "private open data" that is relevant for the efficient operation of the market and in public policy of public interest. •Create collective rights concerning data to bridge the gap between analyzing hidden trends and behavior and their effects on societies and individuals. Identify the technical and "public good" challenge, as AI technologies need high-quality data for training, testing, and interactive dynamic testbeds and simulation environments. •Enable (data) actors that already hold valuable datasets and resources while simultaneously observing commercial and individual rights and interests in the data. •Manage data as strategic national assets to secure access to high-quality datasets. Specify the government's approach to unlocking the power of data through the national data strategy, including access to quality and representative data for robust and effective AI systems. •Assign Central Digital and Data Office to consolidate core policy and strategies for data foundations and work with expert cross-sector partners to improve the government's use and reuse of data to support data-driven innovation in the public sector. •Adopt effective policies and programs and improve service delivery and research outcomes to enhance the government's use of data to benefit the citizens.

**Theme-2: Release data to support leadership in the field** and prevent critical sectors from becoming overtaken by foreign stakeholders [France, UK, the Netherlands, Japan, Singapore, USA]. Develop and implement standards, tests, and measurement methods to make AI tech more reliable, usable, and interoperable, making decisions based on models built using data. •Improve data foundations in the private and third sectors through the National AI R&I program and Open Government License to lead best practices in FAIR (findable, accessible, interoperable, and reusable) data while growing the intellectual and Infrastructure capacities needed to make the best use of the healthy national data ecosystem for accelerating AI development and adoption pathways. •Ensure knowledge, tools, and training modules available for the responsible use of data. Apply the internationally accepted and verifiable FAIR principles to ensure the standards, tools, and training per domain or sector. Address the urgent need to increase the speed of growth in the effective use of big data, knowledge, and computing resources and increase AI applications contributing to social implementation. •Use data across its lifecycle, from problem statement definition to exploitation. •Address the need for vetted and openly available datasets with identified provenance to enable reproducibility and confident advancement in data-intensive science.

**Theme-3: Address the silo effect** from the lack of forward-looking and cross-cutting thinking, which leads to isolated prioritization of systems incompatible with future AI developments. Silo logic also includes fear of losing control, which hampers the circulation of data [France].

**Theme-4: Review the privacy act** to ensure that it is suitable in the context of more personal information about individuals being captured and processed and empower consumers to protect their data [Australia, France, USA, UK, Netherlands]. For example, Australian Data Strategy specifies how the government will enhance the effective, safe, and secure use of data and provide a greater understanding of government policy. •Users need appropriate tools and services to manage technical relations and transfer personal data from one service to another. •Make privacy methods sufficiently explainable and transparent to help researchers make them safe, efficient, and accurate and remain mindful of dangers from data access. •Explore privacy-enhancing technologies that can remove barriers to data sharing by managing risks from sharing commercially sensitive and personal data and new directives to explicitly allow the collection and processing of sensitive and protected data to monitor and mitigate bias in AI systems. • Collect large-scale data to give companies more information about consumer behavior and the most effective nudges to change that behavior.

**Theme-4: Establish a democratic data characteristics balance** between personal implications and the public good through ethical and legal standards [Spain, France]. •AI improves transparency and public activity disclosure, using applications adapted to individual needs, thus enabling technology to enhance the quality of life. For example, increase speed and accuracy of historical personal health data processing for diagnosis, allowing early detection of anomalies. •Launch and accelerate "data for social good project" to stimulate the use and good governance of public and private data. It will generate social and public benefits through open data, citizen-generated data, and government-to-citizen data to streamline processes to enhance the quantity and quality of public participation in government.

**Theme-5: Develop mechanisms to ease the government's adoption of AI systems** with greater emphasis on data privacy, security, and other specific concerns [USA, Singapore, Canada, Spain, Germany, Netherlands]. For example, the formation of a task force to perform a horizontal scan across government agencies to find application areas within agencies and determine the concerns to be addressed for enabling adoption •Modify data acquisition processes across agencies to incorporate specific requirements for AI standards in 'request for proposal' and encourage communities to engage in standards development and adoption, including good practices that foster healthy competition between tech developers, such as community-based benchmarking (like TREC) and lower barriers and strengthen incentives to make training and testing data accessible. •Establish digital and data infrastructure to enable national transformation drives for government, society, and the economy. •Establish reliable data flow, Infrastructure, pipelines, and structured and unstructured data storage. For example, data-driven insights for teachers to design learning. •Initiate high-performance, efficient, secure, trustworthy, and sovereign public-private federated data infrastructure to fully enable cloud computing providers to comply with European quality and performance standards (like GAIA-X). •Ensure computing power and (local or remote) data processing necessary for practical AI applications. •Compare greater infrastructural requirements for AI services with cloud/software-as-a-service systems due to the need for accessing expensive high-performance computing and large datasets for developing and deploying some AI models.

**Theme-6: Combining data from an effective coordinated collection by different parties** will enable AI applications to contribute toward better work processes within government organizations and better societal solutions [Netherlands, Italy]. •The availability of usable and shareable high-quality data is crucial for developing high-quality and reliable AI applications, such as assisting car manufacturers in identifying faulty cars and doctors in making better diagnoses. High-quality data is measured by accuracy, completeness, representativeness, timeliness, verifiability, and risk of bias. •Increase the quality and lower the cost of customer services through a higher level of personalization.

**Theme-7: Develop techniques for visualization** and human-AI interfaces [USA]. Better visualization and user interfaces are additional areas that need much greater development to help humans understand large-volume modern datasets and information from various sources. Visualization and user interfaces must present increasingly complex data and information derived from them in a human-understandable way

### *3.4.2. Data for Sustainability Goals*

**Theme-8: Develop AI-enabled ecological solutions** using open public data through public-private partnerships for data on weather, agriculture, transport, energy, biodiversity, climate, waste, land registry, and energy performance assessments [France, Netherlands, France, Japan, Korea, Germany, Spain]. •Provide access to sensitive data for specific applications to address sectoral challenges. For example, auxiliary power unit data can make flight operations more reliable and lower maintenance costs. •Release the data for ecological transitions will enable AI to contribute toward the greening process. Deep mind-enabled energy consumption in data and cooling systems. AI can support the intelligent ecological transition to championing sustainable and ecological ecosystems enabled by academic ecosystems, research, and a wealth of data.

AI can process and model ecological data from high-speed biodiversity and water quality imaging sequences. •Systematically analyze energy consumption and environmental footprint to achieve energy-efficient AI activities that benefit the planet across all sectors. •Collect and analyze data on energy consumption to promote energy-saving behavior, such as intelligent personal nudges. •Fine dust forecasting complements real-time monitoring of underground water pollution by livestock manure. •Leverage efficiency potentials for data-driven Sustainable Development Goals and strengthen AI applications in the environmental sector, such as climate impact assessment, ecosystem analysis, mobility analysis, and energy consumption behavior. Increased funding for satellite earth observation will encourage environmentally friendly urban developments, transport, and mobility and manage the sustainable use of agriculture, forestry, raw material extraction, water, and energy. •Linking data for logistics and supply chain efficiency improvements. •Use Data and AI as the basis of logistics operations, such as the commercialization of space transport.

*3.4.3. Data Collaborations and Ecosystem for Research, Development, and Innovation*

**Theme-9: Renew equipment to improve performance** to keep up with data growth [France, Germany]. New computer architecture, with more efficient use of cloud computing, and highly energy-efficient computing and memory, such as neuromorphic systems, can reduce the need for centralized data but increase the demand for equipment renewal. In addition, larger computing and storage spaces provide better energy efficiency than smaller data centers. •Establishing a national supercomputing center considers resource efficiency and industry use possibilities while anticipating future peak demand for AI applications and large data volumes analysis. s

**Theme-10: Research to improve the efficiency of data techniques** to address the integrity, improve data quality, and a new way to extract data and metadata simultaneously [USA]. •It includes methods and approaches for data cleaning, knowledge discovery, discovering anomalies and inconsistencies, and incorporating human feedback. •Research to enable machine learning algorithms to learn from high-velocity data, including distributed ML that learn simultaneously from multiple data pipelines, and for the AI systems to intelligently prioritize and sample data from large-scale simulations, experimental instruments, and distributed sensor systems

**Theme-11: Apply machine learning to encrypted data** with high throughput and accuracy [USA, France, Spain]. •Advancing hardware (such as sensors) and algorithms to 1) enable robust and reliable perception, capturing data at longer distances, higher resolution, and in real-time; 2) improve data-intensive AI methods; and 3) responsive controlled data pipelines throughout distributed systems for immediate data availability. These will enable AI development to create new sectors of the economy and revitalize industries. Adopting AI's fundamental technologies can drive profound economic impact and quality of life improvements. •The tightly interwoven Infrastructure of emerging technologies such as 5G for big data and the strengths of universities to enable high-performance computing, cybersecurity, and digital-enabling technologies will address intense digitalization and technological developments. Hardware such as GPU (Graphics Processing Unit) technology enables efficient AI techniques to use increasing amounts of large data. •DARPA launched a multiyear investment in the "AI Next" campaign to improve the robustness and reliability of AI systems, enhance the security and resilience of machine learning tech, reduce power, data, and performance inefficiencies, and pioneering next-generation explainable AI algorithms and applications. Foundational investments to drive machine learning includes data provenance and quality, novel software and hardware paradigms, and the security of AI systems.

**Theme-12: Form situational awareness through integrated data from perception**, including geometric, location, and velocity attributes [USA, Singapore, Japan, France]. This will provide AI systems with 1) comprehensive knowledge and 2) a model of the world to plan and execute tasks effectively and safely. •Use AI (natural language processing and machine learning) to transform unstructured data into reliable, contextualized data sources for corporations. •Build a platform that combines satellite and ground data to perform complex AI analysis. •Extract knowledge from texts or databases according to criteria of novelty or similarity using text and data mining. •Establish data Ecosystem (enabled by partnership and funding) for AI research, development, and innovations.

**Theme-13: Establish interactions between academia and the private sector and government funding** to enable innovations to improve accuracy in AI models, such as reduced error rate in speech recognition and enabled real-time translation [USA].

**Theme-14: Create the right conditions for high-quality and leading AI research and innovation with excellent training, talents** for living and working with AI, and more usable data for enabling AI application developments [Netherlands, UK]. •Enable high-quality digital and intelligent connectivity, measured by latency, data rate, and reliability. Specific applications, such as self-driving cars, need significantly large high-quality data and hence require high reliability and low latency digital connectivity. Likewise, Augmented and Virtual Reality applications require a high data rate and low latency. •Skills, data, and infrastructure investments will enable AI study and cutting-edge research.

**Theme-15: Facilitate collaborations between firms and scale commercialization** by driving innovation and the digital economy's governance using common, adaptable frameworks, such as industry standards and conformity assessments, to provide clear guidelines and best practices from internal data governance and analytics practices [Canada, UK].•Provide supports for affordable Intellectual Property (IP) advice or centralized IP education and digital infrastructure, including legal and technical infrastructure for data sharing, affordable broadband, data storage and processing, and high-performance computing. •Work across government agencies and businesses to tackle complex undefined problems and explore legacy data use to enable the translation of fundamental scientific discoveries into real-world applications.

**Theme-16: Establish the basis of the AI ecosystem on secure and sovereign data infrastructure** to ensure competitiveness while skilling citizens to deal with AI applications confidently in various situations [Germany, France, Netherlands]. •Address the problem that although a company with a wealth of data history becomes a potential purchaser of AI solutions, it may need help in 1) formalizing its requirements and 2) identifying stakeholders that can provide them with solutions. •IP rights to AI systems are complex due to the components. For example, it is increasingly difficult to distinguish between data and algorithms, as both are often called "code." •Once a platform achieves a powerful position, its self-reinforcing processes can make it challenging for SMEs and startups to catch up.

**Theme-17: Attain international competitiveness** through next-generation AI technology, R&D, and deployment, and form openly compatible, stable, and mature technological systems [China]. Research, establish, and operate vehicle automatic driving and road coordination technology systems. Research and develop information and integrated data platforms for transportation under complex multi-dimensional conditions.

#### *3.4.4. Data Security and Protection*

**Theme-18: Guarantee personal data protection** through legally robust agreement and sufficiently robust framework for businesses [France, Netherlands]. • Legislation on data protection only regulates AI algorithms, as they are primarily based on personal data, and their results affect individuals directly. •There are many efforts to standardize data in the healthcare sector, such as the Basic Data Set for Care initiative.

**Theme-19: Collect high-quality data and keep it safe** from cyber-attacks and other risks, including innovative ways of safeguarding data while facilitating practical usage and accelerating data-driven research [Japan, Singapore, USA]. •High-quality data is enabled by a smarter research environment and a data platform to accumulate them. •Identify how data pre-processing and additional analysis impact data quality.

**Theme-20**: **Launch the National Institute of Cybersecurity** to promote companies' cybersecurity in collaboration with European agencies, aiming to guarantee data security and effectively coordinate all agents and administrations, including regional authorities [Spain].

**Theme-21: Use AI applications to strengthen fraud prevention systems** by adopting mechanisms for detecting suspicious behavior while analyzing data and documents [Italy, China]. •Data and security detection platforms should support AI for public data resource libraries, standard datasets, test evaluation methods, techniques, norms, and tools, promoting open-sourcing and openness of software and technology cloud service platforms.

**Theme-22: Address the complex or confusing task of navigating and applying data protection provisions** as they may impede the uptake of AI and for organizations to develop or deploy AI systems [UK, USA]. For example, Civil groups raised objections around the known vulnerabilities in mass surveillance for malicious purposes and social media sites being scrapped for use that violates the terms and policies for data use.

### *3.4.5. Data Quality and its Fair and Safe Use*

**Theme-23: Develop new metrics for data** performance, interoperability, usability, safety (prevention against harm), privacy, confidentiality, traceability, and fairness (against potential bias) [France, India, Spain, USA, Canada, Germany]. •Policymakers rely on standards (as agile regulatory instruments) to keep pace with innovation, evolving data practices, and societal expectations. •Under GDPR, preserving confidentiality, integrity, and resilience of the processing systems and services requires the best encryption practices or coarsen data resolution to mask and de-identify personal/sensitive data. • The private sector safely develops AI solutions, products, and services while applying best practices for security and data quality. For the public sector, optimize the quality of analysts-based public policies using simulations. •Develop and standardize testing methodologies and metrics, provide objective data to benchmark, and effectively track and evaluate the advancements aimed at strategic scenarios. Increasing AI dataset availability is essential for researchers to use actual operational data to model and run experiments on real-world systems and scenarios. For example, NIST developed comprehensive standard test methods and the associated performance metrics to statistically assess emergency response robots' critical capabilities. •The government will foster data quality assurance through benchmark tests, reference data, and setting up the data for validating algorithms to ensure reliable AI methods. •Design tasks, incentive structures, and feedback mechanisms to ensure accurate and unbiased annotation.

**Theme-24: Check fairness and equity** by testing the large quantity of the system’s user profiles against various false input data according to precise guidelines, such as checking the gender equity of a recruitment website. For example, auditors can obtain accurate data on male-female discrimination in the workplace [France].

**Theme-25: Implement data quality verification and assurance** and propose an international standard for AI life cycle and quality assurance that includes data [Japan, USA, Korea, Germany, UK]. •Prioritize improvements to access and quality of AI data and models from the AI research community's feedback. Inherit AI vulnerabilities and performance requirements when AI components are connected to other systems and information safely and securely. •Follow applicable laws and regulations, carry out dataset development, and ethically share while recognizing risks associated with inappropriate use, inaccurate or inappropriate disclosure, and limitations in data de-identification techniques (hence privacy and confidentiality are not ensured). •Establish standard data for clinical verification and a professional review system to improve the quality of AI medical devices and speed up the commercialization process. •The government must strengthen science-based quality criteria for medical study designs, bolster research funding on specific AI in healthcare challenges, and develop the infrastructure of universities. The infrastructure will ensure quick and wide availability of standardized, high-quality health-relevant data that enable AI-driven health research. •One of the largest centralized collections of medical images in the UK (NCCID) is being used to study and understand the COVID-19 illness and improve the care of hospitalized patients with severe infection.

**Theme-28: Ensure future funding and programs incentivize cross-projects for sustainable utilization of results**. Projects must adopt open-source algorithms, best practices, and quality assurance to ensure research findings' traceability and verifiability [Germany, USA, France, Netherlands, India]. •Address the rapid growth in scientific and societal understanding of AI security and safety. Otherwise, AI systems can be made to do, learn, and reveal the wrong thing through adversarial examples, data poisoning, and model inversion. •Address the difficult, time-consuming, and costly process (for both human and financial resources) to enhance the annotation of datasets to realize the importance and potential benefits fully. •Address the issue with newly acquired data sources, resulting in a more complete, complex, and possibly ambiguous picture. •Address the issue of how AI systems can appear to have prejudices amplified on large-scale applications. For example, healthcare discrimination towards certain minorities, due to the black-box nature and self-learning ability, makes it hard to justify decisions and apportion liability for errors. •Address the issue with deep learning for personalization and decision-making, which can have social inequalities embedded in the decision algorithms

#### *3.4.6. Data Skills and Education*

**Theme-29: Establishing an institute for continuing professional development** for businesses and researchers across disciplines, including hiring postdoctoral researchers, is as essential as traditional bachelor/master education for sharing data, experience, and best practices for validating AI solutions [France, Singapore]. •Integrate talent, data, regulation, and effective deployments across departments to fully realize the positive impact of AI.

**Theme-30: Address gender balance** of the science and tech sectors' workforce, which can result in unconscious bias in algorithms [France, Netherlands, Canada]. •Grow both male and female AI experts and data professionals to support the rapid growth of AI and computer science programs. Lack of (gender) diversity can result in algorithms that produce cognitive biases in the system design, data analysis, and interpretation of results. •Investment in AI research (the core focus of AI strategy) needs to be supported by a strong, equitable, and accountable data-centric talent pipeline to commercialize and govern responsible innovation and sustainable, inclusive growth of industrial AI.

**Theme-31: Strengthen students' academic qualifications with AI expertise** and improve the quality and performance of higher education through responsible use of AI, fostering innovation through big data and AI-enabled development of university courses and learning design settings [Germany, Netherlands]. •Address how AI and computer science education programs are growing so rapidly that it is difficult to accommodate the growth.

### 3.5 Analysis of Strategic Themes

Based on a thematic analysis of Leximancer's generated passages on "data and related concepts," a set of strategic themes were derived. These themes have been checked to be consistent with the executive summaries and can encompass (overarching) all the key concepts within the data-focused AI strategies, as discussed in Section 3.4. Given that there were 3414 passages, the key themes of key action items could be identified within the first 150 passages, after noting some of the passages of #121 onwards were becoming quite repetitive. Therefore, we refined the initial structure and thematic list of action items based on the next passages but stopped at passage #221 after no further refinement was needed.

This section will detail the **Strategic Themes** from the leading nations, forming the future mapping of AI strategy:

**E**: ***Education*** and training reform, ensuring jobs and skills, anticipating future demands and events, **promoting in-depth AI knowledge**, bringing researchers into teaching, and promoting streamlined career transfers between academia and private sectors.

**P**: ***Partnerships*** for R&D&I collaboration between government, industry, academia, and a society characterized by **interdisciplinary research and data sharing plans**, increased knowledge generation, translational outcomes, and quality deployment.

**I**: ***Infrastructure*** of technology, people, and policy for critical management of data and AI methods that bolsters **the trustworthy application of AI** and promotes *zero-cost usage*, leveling the playing field for SMEs and researchers.

**C**: ***Community*** impact from AI innovations that promote the public good, including ecological agenda, energy efficiency for storage, and green computing, promoting a technological future underlined by **sustainability-focused AI maturity**, including Infrastructure, collaborations, and education.

As shown in Figure 3, the pyramid of "**EPIC**" strategic themes signifies that the foundation of AI maturity starts with an education that enables meaningful partnership, which is critical to building the infrastructure for quality deployment. At the top of the pyramid, community-driven AI innovation is the capstone for AI maturity. There is an emerging call for increased "AI for good" that promotes social, sustainability, and ecological agenda.

The first fundamental and underlying requirement for AI adoption is literacy, skills, and expertise for data, AI, and related technologies through education, training, and reform. If society is unaware of AI capabilities, they are less likely to be receptive to AI adoption and support new applications and solutions deployments. Therefore, a sustained pipeline of skills and expertise to develop AI is critical for enabling the next strategies. The next layer strategy (on top of education) is partnership and cooperation between skilled players to advance AI applications' technological readiness level for quality deployment via translational outcomes of cutting-edge research, development, and innovation process underpinned by interdisciplinary collaborations. The next layer strategy (on top of the partnership) is establishing infrastructure for data and AI, which is essential to support trustworthy AI application developments driven by the partnership's stakeholders' common values, mission, and objectives. Hence, a partnership is a critical requirement for motivating meaningful collaborative efforts in bringing together the infrastructure of people, technology, and policy, as every player from academia, industry, government, and society has a key role to play in bringing these components together and build the ecosystem and trusted network. Finally, community-driven AI innovation provides the overarching principle and mission for sustaining and developing AI maturity, supported through infrastructure, collaborations, and education.

These strategic themes have a direct mapping (one-to-one correlation) with findings from existing reviews government's role in enabling the efficacy of the National AI Ecosystem (as previously depicted in Figure 2). E => Prepare the nation's workforce for the new era of AI via education and training opportunities. P => Prioritize AI research and development investments, diffusing AI tech across all industry areas. I => Enhance AI infrastructure and access to high-quality cyberinfrastructure and data. C => Develop technical standards for AI (which are community impact oriented, such as green computing).

Table 5 lists each theme's specific actions, expected outcomes, and benefits. This analysis demonstrates that the EPIC strategic themes can cut across and potentially solve the barriers and challenges for AI adoption. Each theme encompasses and can solve AI's political, economic, societal, technological, and environmental challenges.

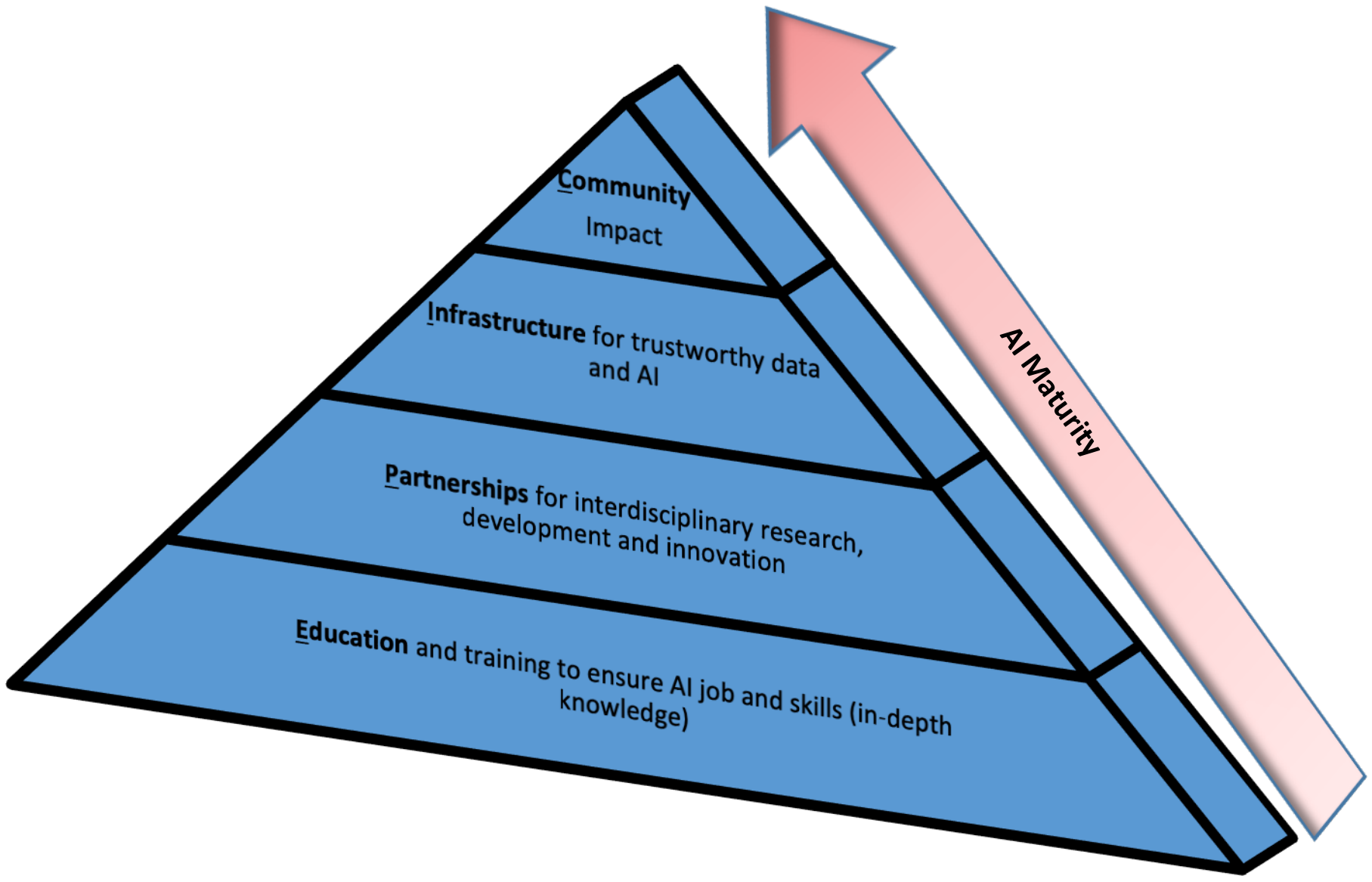

Figure 3. "EPIC" Strategic Themes of Data and related concepts (the pyramid indicates the dependency from fundamental to the high-level strategic theme, whereby education is fundamental to building partnership, which is critical to driving the infrastructure, etc.)

Table 5. EPIC Strategic Themes from leading nations' AI strategies

| | **Strategy** | **Expected Outcomes and Benefits** |
|---|---|---|
| **E1** | Overhaul **formal education and lifelong learning** to develop creative skills in AI graduates, which may involve a more technical in-depth 3-year vocational/bachelor's degree for AI. | Strengthen the current and next-generation personnel's skills and capacity to develop creative AI applications. |
| **E2** | Develop **sector plans** for applied sciences, training, and connection with the labor market. | Promote a strong supply-demand for AI scientists and researchers. |
| **E3** | **Embed research** into curriculum due to the rapidly changing sector of AI. | Strengthen and increase the appeal of AI education through up-to-date and evolving knowledge from research. |
| **E4** | **Integrate researchers** into higher-degree education, industry, or government, which includes industry advisors for doctoral study. | Promoting knowledge sharing and applied research outcomes during research education, training, and practical learning. |
| **E5** | Promote greater flexibility in attracting, retraining, selecting, and **planning personnel** in service provision from cross-disciplinary, industry, and sector backgrounds | Promote easier (more flexible) transition for labor/skills in different service sectors. |
| **P1** | Promote **research as the critical driver** for discovery and growth, which includes discovering new AI methods and algorithms. | Generate new knowledge, technologies, and disruptive innovations. |
| **P2** | Promote **networking synergies in the initial phase**, including hackathons, and reward excellence and formation of startups and spinoffs in academics and business. | Align the different speeds of AI tech implementation according to the sector, the strategic nature, and the degree of maturity and development. |
| **P3** | Enable **AI innovation ecosystem** with a vibrant partnership condition that supports AI R&D, marketing, and deployment of AI innovations. | Stabilizing R&D&I to enable knowledge transfer, promote strong translational outcomes, streamline research products, and stabilize personnel. |
| **P4** | Enable **interdisciplinary team**s working together within public-private partnerships on prioritized sectors based on society's requirements. | Promote fast and targeted/focused outcomes and impacts, avoid spreading efforts too thinly, and drive specific application areas and sectors. |
| **P5** | Plan for multi-channel and comprehensive input for AI **funding and grants** by the government and market. | Establish marketized for continuity and impact of AI developments. |
| **P6** | Promote **trusted cooperative networks** for a streamlined AI innovation process, incl. the environment and governance. | Increasing the feasibility of AI use in government and across applications/sectors. |
| **P7** | Increase **competitiveness and visibility** of national AI knowledge, expertise, and infrastructure; and create openly compatible, stable, and mature technological systems within the international region. | Enable the highest quality world-class AI R&D and applications deployment that promotes best practices. |
| **P8** | Innovative **contract engineering**, including formalizing risk-taking, does not penalize failure. | Incentivize data and AI capabilities sharing, protect purchasers/adopters, and manage risks for R&D partnerships. |
| **I1** | Establish sector-specific **regulations** on the use of algorithms (such as for trading) and region-specific (e.g., European) regulations on platforms and businesses. | Align AI algorithms and applications with sector-specific suitability and expand the boundaries of the potential impacts to the surrounding nations. |
| **I2** | Align the **supply-demand** of public data with the needs of AI applications, ensure policy for streamlining the process of collecting, using, and reusing data, remove obstacles from opening access to public data, and specify terms and conditions for accessing private data. | Promote public access to open data and regulate private data sharing, which subsequently incentivizes cross-project and open-source algorithms. |
| **I4** | Ensure **traceability and verifiability** of AI research findings and project outcomes evaluation. Increase access to and the required expertise to assess training data. | Sustaining utilization of AI research outcomes through verified results and addressing risks of discriminatory AI applications due to biased state-of-the-art AI algorithms. |
| **I5** | Ensure **ethical and accountable AI** throughout its development and use AI methods for preventing discrimination during decisions making. | Tackling biases that can be caused by the way the AI solution is developed (beyond research), including the AI methods, algorithms, and training data. |
| **I6** | Promote **digitalization** of society, business, and government and create contexts for compiling, processing, and analyzing data. | Maintain a realistic understanding of the strategic values to prosper with AI and data. |
| **C1** | Promote the use of **AI for sustainability**, such as green computing for training AI models and methods. | Enable energy-efficient AI activities that are good for the planet. |
| **C2** | Adopt a paradigm shift in sharing and collaboration for incorporating **ecological agenda**. | Support frugal models of tech and economics toward energy-efficient collective growth. |
| **C3** | Enable startups and small to medium enterprises (SMEs) to **work with established platforms** for energy-efficient and scalable storage and computing. | Enable startups and SMEs to leverage established platforms, which would take much work to compete or catch up with, and benefit from the significant potential for growth and self-improvement. |

## 4 DISCUSSIONS AND FUTURE WORK

The **E-P-I-C strategic themes** can guide the development of AI strategy in every organization. **E**ducation is the basis for ensuring a strong understanding of the latest socio-technical fundamentals, including the practical knowledge and skills for taking full advantage of AI technologies. Being "AI educated" enables the instigator of AI activity to start with a fully informed decision about whether the organization has achieved or completed the objectives from their existing AI deployment and the enabling ecosystem. AI deployment results can range from using AI to help extract useful information from data to fully automating actions and decisions. A completed AI activity generally produce AI application (products and services) and Infrastructure. AI infrastructure includes technology (data, AI models and algorithms, and digital connectivity), people (skills and workforce plan), processes, and workflow (e.g., data collection, storage, and use/reuse). Moreover, Infrastructure includes policies and regulations to ensure collective benefits and protect individual interests, such as privacy.

The first strategy is to assess the AI capability and readiness, including data, digital connectivity, and AI methods and algorithms as the key components. Then, depending on the AI capacity and readiness, the organization can determine the need to start or iterate AI activity, either AI deployment or AI ecosystem enablement (or both simultaneously). For both goals, EPIC strategic themes can help set target goals and guide the process from the start until the end.

Community strategic themes guide strategies for the successful development and deployment of AI applications and drive the planning and designing of the ecosystem to develop responsible AI, including the methods for aligning the requirements to realize social impacts fully. The **I**nfrastructure's strategic theme can help drive the decisions and actions around the design and implementation of the necessary process, technology, and people infrastructure to support the AI application and monitor the responsible AI checklist throughout the process. The **P**artnership strategic theme helps to bring together the resources, including combined finance and established protocol for engagement and procurement.

Each actor from academia (researchers), industry (business manager and private sector leaders of the industry), government (minister and public services of the industry), and society (end users and entrepreneurs) has a distributed leadership role to co-govern the development and deployment process. They must collaborate to co-create the AI application or its components iteratively through research, development, and innovation collaboration. The byproduct of partnership is a trusted network of people, processes, and technology infrastructure for AI deployment, which encourages adaptive and hybrid governance models that evolve with the AI maturity of every player. Finally, the **E**ducation strategic theme can drive the process of operating, monitoring, and scaling while considering the adoption barriers, which include training the end-users and business operators and informing the wider community about AI application's use and benefits. Education is also the basis of designing skills and employment pathways for next generation AI specialists, business managers, and end users.

Let us start with the end in mind to develop strategies for enabling the AI ecosystem. First, the **C**ommunity strategic theme drives "data for sustainability goals" strategies to manage the development of technical standards, such as verifiable AI models and energy-efficient algorithms, which can contribute toward the ecological agenda. Second, the **I**nfrastructure strategic theme uses "data quality and fair use" and "data security and protection" strategies to direct infrastructure enhancements, such as secure data release, data architecture, and software platforms for research data. Third, the Strategic partnership theme drives "data collaborations and ecosystem" strategies to administer the R&D investments, such as strengthening national research structures and focusing AI research efforts into priority areas. Finally, the **E**ducation strategic theme drives "data skills and education" strategies to manage and prepare for the workforce, such as regulation for work settings and product safety, and networking and international cooperation to increase collaboration and access to knowledge.

**Recommended future work**: by conducting further reviews and analysis of national AI strategic documents from developing countries, the future mapping AI strategy can be adjusted or extended to incorporate specific requirements for these countries and ensure the goal of "leaving no one behind." Furthermore, by applying the strategies to specific AI applications, industries, and the public sector, further details and considerations can be added to help tailor the recommendations and strategies toward the specific requirements and demonstrate the robustness and applicability of the strategies. In addition, the strategies should incorporate reviews and analysis of different expertise. For example, future strategies must incorporate leading researchers, consulting firms, AI businesses, and entrepreneurs, as well as lessons from longitudinal and long-term case studies of applying the strategies in real-life situations. Such multi-perspectives strategies can formulate further practical enhancements to ensure that the strategies can truly contribute toward promoting long-term benefits from the sustained use of AI. Moreover, there are opportunities for AI strategies to inform and influence the complementary regulatory guidelines for enabling co-governance of Responsible AI implementation in specific domains and industry sectors.

## 5 CONCLUSIONS

People's concerns for AI and its potential are the future of work, losing jobs due to automation, lack of understanding or bloated expectations from AI, and the need to trust AI systems with personal data. Security and privacy issues are becoming more prevalent due to the rapid growth of technology, business/organizational transformation, and increasing dependency on digital connectivity and the use of embedded AI. Increased interests in global socio-economic and environmental issues and challenges have brought the convergence of aligned individual and organizational requirements and constraints (i.e., the triple bottom line) and raised opportunities for shared values, motivation, and objectives of AI adoption. Co-governance throughout AI deployment will promote leadership through principles, standards, and regulations that evolve through AI maturity.

Everyone has a role to play in advancing AI for the future, with next-generation algorithms and applications increasingly needing to consider long-term impacts through multi-perspectives strategic themes. AI maturity must start with lifelong learning, transformational education, and skills for community awareness, human resources, AI literacy and competency, and AI development leadership. Through a quadruple partnership by society, government, businesses, and academia, AI capacity, readiness, and competitiveness (locally and globally) can be achieved through collective efforts and a trusted network underpinned by shared interests and resources, including funding, AI models, algorithms, and datasets. The partnership and cooperation will help to ensure and enhance the equitable establishment of the quality infrastructure of the AI ecosystem, including people, processes, and technology, that promotes rapid research, development, and innovation to develop and deploy novel, trustworthy AI applications. The capstone of AI maturity focuses on community impact, including sustainability goals development through ecological AI applications, greener AI computing, responsible development, deployment, and scaling of AI innovations. The EPIC strategic themes from the leading nations can guide future mapping AI strategy, as the current and emerging strategies for AI application deployment and AI ecosystem strengthening can be expanded through the lens of education, partnership, infrastructure, and community impact. The underpinning requirement for enabling AI capacity and maturity is education and employment pathways that include AI to augment the future workforce and jobs, as there is always a space for human intelligence and involvement in the shared responsibility of maintaining responsible adoption and long-term use of AI.